\documentclass[12pt,preprint]{aastex}
\usepackage{epsfig}
\usepackage{emulateapj5}
\begin{document}

\title{A Young Planet Search in Visible and IR Light: DN Tau, V836 Tau, and V827 Tau$^1$}

\author{L. Prato\altaffilmark{2,3}, M. Huerta\altaffilmark{4},
C. M. Johns--Krull\altaffilmark{3,5}, N. Mahmud\altaffilmark{3,5},
D. T. Jaffe\altaffilmark{6}, P. Hartigan\altaffilmark{5}}

\altaffiltext{1}{This paper includes data taken at The McDonald Observatory
of The University of Texas at Austin.}
\altaffiltext{2}{Lowell Observatory, 1400 West Mars Hill
  Rd. Flagstaff, AZ 86001; lprato@lowell.edu}
\altaffiltext{3}{Visiting Astronomer at the Infrared Telescope Facility,
which is operated by the University of Hawaii under cooperative agreement
NCC 5-538 with the National Aeronautics and Space Administration, Office of
Space Science, Planetary Astronomy Program.}
\altaffiltext{4}{American Astronomical Society, 2000 Florida Avenue Suite 400,
Washington, DC 20009; marcos.huerta@aas.org}
\altaffiltext{5}{Department of Physics and Astronomy, Rice University,
MS-108, 6100 Main Street, Houston, TX 77005; cmj@rice.edu, naved@rice.edu, hartigan@rice.edu}
\altaffiltext{6}{Department of Astronomy, University of Texas, R. L. Moore
Hall, Austin, TX 78712; dtj@astro.as.utexas.edu}

\begin{abstract}

In searches for low-mass companions to late-type stars,
correlation between radial velocity variations and
line bisector slope changes indicates contamination
by large starspots.  Two young stars
demonstrate that this test is not sufficient to rule
out starspots as a cause of radial velocity variations.
As part of our survey for
substellar companions to T Tauri stars, we identified the
$\sim$2 Myr old planet host candidates DN Tau
and V836 Tau.  In both cases, visible light radial velocity modulation
appears periodic and is uncorrelated with line bisector span variations,
suggesting close companions of several M$_{Jup}$ in these
systems.  However, high-resolution, infrared spectroscopy
shows that starspots cause the radial velocity
variations. We also report unambiguous results for V827 Tau,
identified as a spotted star on the basis of both visible light
and infrared spectroscopy.  Our results suggest that infrared follow up
observations are critical for determining the source of radial
velocity modulation in young, spotted stars.

\end{abstract}


\section{Introduction}

Extrasolar planets are common; over 300 systems have
been discovered (e.g., exoplanet.eu).
Recent studies have targeted
higher mass \citep{sat07, joh07}, lower mass \citep{but06, end06},
and younger objects \citep{pau06, set07}.  
Identifying young planets is important to define the time scale for
planet formation and thus distinguish the possible formation process(es).

Young stars still surrounded by the
circumstellar material which forms planets
are typically located at distances of $>$100~pc and are thus
inherently faint and often obscured.
They also manifest strong magnetic activity \citep[e.g.,][]{cmj07}
and are highly spotted.  Numerous, large spots complicate detection of
extrasolar planets through radial velocity (RV) monitoring \citep{saa97}
because a spot that is partially visible at all times on the
surface of an inclined star mimics RV modulation
\citep[e.g.,][]{bou07, hue08}.

\citet{pau06} studied 12$-$300~Myr old nearby
stars and found no evidence for planets
with masses $>$1$-$2~M$_{Jup}$ at the 3~$\sigma$ level.
\citet{set07} identified a minimum mass 6.1~M$_{Jup}$ planet in
a 852~day period orbit around the 100~Myr old G1$-$G1.5~V star HD~70573.
More recently, \citet{set08} reported a $\sim$10~M$_{Jup}$ planet 
in a 3.56 day orbit around the 10~Myr old star TW Hya, although
\citet{huel08} identify this result as attributable to spots.

In this letter we present our observations of the young stars
DN Tau, V836 Tau, and V827 Tau.  While the V827 Tau visible light data
clearly implicate spots as the cause of the apparent RV
variability, corresponding data for DN Tau and V836 Tau suggest the
presence of giant planets.  Our infrared (IR) observations show
that spots cause the RV variations seen in all
three stars.  In \S 2 we describe the
observations, in \S 3 present our data analysis and 
the evidence for spots, and in \S 4 provide a brief discussion.
We summarize in \S 5.

\section{Observations and Data Reduction}

\subsection{Visible Light Spectroscopy}

Visible light spectra of DN Tau (M0), V836 Tau (K7), and V827 Tau (K7)
were taken at the McDonald Obervatory 2.7~meter
Harlan J. Smith telescope, between November, 2004 and January, 2008,
with the Coud\'e echelle spectrograph
\citep{tul95}.  A 1.2$''$ slit yielded R$\sim$60,000.
Integration times were $\sim$1800~s; average seeing
was $\sim$2$''$.  ThAr exposures taken immediately before and after
each spectrum provided wavelength calibration; typical
RMS values for the dispersion solution precision were $\sim$4~m~s$^{-1}$.
RV standards \citep{nid02, but96, cum99} were observed on
every night of every run; their overall RMS scatter
is 140~m~s$^{-1}$.  We obtained 43 spectra of DN Tau,
21 of V836 Tau, and 20 of
V827 Tau and applied standard IRAF reduction routines.
Details are given in \citet{hue07} and \citet{hue08}.

\subsection{Infrared Spectroscopy}

We observed DN Tau, V836 Tau, V827 Tau, and the RV
standards HD 65277 (K5) and
GJ~281 (M0) on UT 2008 February 13$-$20 with CSHELL \citep{tok90, gre93},
the high-resolution, IR  spectrograph at the NASA IRTF 3-m telescope.
The seeing was 0.4$-$0.8$''$.  Each object was observed on 6$-$8 nights.
The 0.5$''$ slit yielded R$\sim$46,000.  We obtained data
in 10$''$ nodded pairs.  Spectra were centered at 2.298~$\mu$m (vacuum).
Integration times were $\sim$1 hour for the T~Tauri stars and $\sim$8 minutes
for the standards.  The signal to noise ratio (SNR) was $\sim$70$-$120.
Data were reduced as described in \citet{joh99}.

\section{Analysis}

\subsection{Radial Velocities: Visible Light}

Relative RVs were determined by cross correlating
a high SNR, fiducial spectrum against
all other spectra for the same target.  We used six orders
spanning $\sim$5700~\AA\ to 6800~\AA.  Uncertainties
were estimated from the standard deviation of the mean for the
6 orders, added in quadrature with the 140~m~s$^{-1}$ uncertainty derived
from the RV standards \citep[\S 2;][]{hue08}.  RVs were
corrected for the Earth's barycentric motion.

Optimum periods and uncertainties for phasing the RV data were selected based
on power spectra \citep{hue08}.  For DN Tau we found
P$=$6.33~$\pm$0.20 days and a false alarm probability (FAP) of $< $0.001, for
V836 Tau, P$=$2.48~$\pm$0.49 days and FAP$=$0.10, and for 
V827 Tau, P$=$3.76~$\pm$0.06 days and FAP$<$0.001.  
We also checked for periodicity using the discrete Fourier transform
plus CLEAN method of Roberts et al. (1987).  The strongest power
spectrum peaks for DN Tau and V836 Tau occur at the
same periods.  The CLEAN method recovered a best
period of 3.61 d for V827 Tau, within $\sim$2~$\sigma$
of the above estimate.  The phased RV data
are shown in Figures 1$-$3.

\subsection{Line Bisector Spans}

The presence of a starspot will distort the
line profile at the RV that corresponds to the stellar velocity at the
location of the spot.  This distortion is
in proportion to the ratio of quiescent photosphere surface
brightness and surface brightness within the spot, at the observing
wavelength, and
the fraction of stellar surface covered by the spot \citep[e.g.,][]{que01}.
Thus, asymmetries in the line profiles originating from spots will be
present for all lines in the spectrum of a young star and are typically
correlated with the RV measured from the same spectrum.  These
asymmetries have become the
standard criterion for rejecting starspots as the cause of false RV signals
\citep[e.g.,][]{que01, bou07, hue08, set08}.
For each of the six orders used to determine
the RVs, we cross-correlated all absorption lines and measured the
cross-correlation function (CCF) for that order.  The average of
these six CCFs was used to measure the bisector spans
(lower panels of Figures 1$-$3).  The linear correlation coefficient
and associated FAP \citep{bev92} is listed in the captions.
As expected for a spotted star, a clear correlation between bisector
span and RV is observed for V827 Tau.  DN Tau and V836 Tau show no
correlation, suggesting that the variability is not the result
of spots.

\subsection{Infrared Radial Velocity Signatures}

The contrast between a 4000~K photosphere and a 3000~K spot
\citep{bou89} is greater in visible than in IR light
because flux scales as a steeper function of temperature at
wavelengths shorter than the black body peak
\citep[e.g.,][]{car01}.  Given
the decreased spot to photosphere contrast in the near-IR,
the amplitude of the RV modulation will be smaller.
Conversely, if a planet drives the RV modulation, the
amplitude should be the same in visible and IR light.

\citet{bla07, bla08} used near-IR observations to search for companions to
low-mass objects, exploiting telluric absorption lines for
high-precision RV measurements.  Figure 4
shows an example of a GJ~281 K-band spectrum and illustrates our
similar approach.  We created models by combining high
resolution telluric absorption \citep{liv91} and cool stellar spectra
(the sunspot atlas of Wallace \& Livingston 1992), applying a range 
of velocity shifts relative to the telluric lines.
Other free parameters are $v$sin$i$,
a Gaussian FWHM for the spectrometer line spread function,
scale factors for line depths,
and a first order continuum normalization function.
We employed the Marquardt method for
non-linear least squares fitting \citep{bev92} of
each model to an observed spectrum.  The difference
between the stellar and telluric velocities in the best fit
model spectrum yields the RV,
which was then corrected for barycentric motion.

Figures 1$-$3 show the IR-derived relative RVs.  The
data are phased to the periods
given in \S3.1.  The RV standard deviation in the IR data
for HD~65277 is 127~m~s$^{-1}$ and
for GJ~281 is 98~m~s$^{-1}$.  Internal errors, measured from
the least squares fitting, are $\sim$40~m~s$^{-1}$ for both
standards.  For the young stars, internal errors were
100$-$300~m~s$^{-1}$, depending on the SNR achieved.  Random errors,
which we assume add in quadrature with our internal errors to give
the overall scatter in velocities, are 120~m~s$^{-1}$ for HD~65277 and
90~m~s$^{-1}$ for GJ~281.  We use 110~m~s$^{-1}$ as our final value
for the random errors.  The uncertainties shown for the IR data in
Figures 1$-$3 represent the sum, in quadrature, of the individual
internal error and the 110~m~s$^{-1}$ random error.  Within our
measurement precision, we are unable to detect any IR RV variability in
DN Tau and V836 Tau.  V827 Tau shows significant IR RV
variations but at a reduced amplitude from those observed in
visible light.

\section{Discussion}

Figures 1 and 2 show the visible light RV modulation for DN Tau
and V836 Tau, with full amplitudes of $\sim 1500$ and $\sim 2700$~m~s$^{-1}$,
respectively.  Within the 1~$\sigma$ uncertainties, all but one point in the IR RVs
of DN Tau are consistent with zero; all six IR RVs of V836 Tau also show
no variation.  These results indicate that no planets
are present around DN Tau or V836 Tau with masses greater than a few
M$_{Jup}$ at $<$0.5~AU or $\sim$10~M$_{Jup}$ at $\sim$1~AU, despite the
absence of a correlation between the visible light RVs and bisector spans.
\citet{mat89} identify V836 Tau and V827 Tau as RV variables with 
peak-to-peak amplitudes of 7$-$8~km~s$^{-1}$.  Apparently the density and
size of spots on V836 Tau vary; historical data may
serve as an additional criterion for heavily spotted young stars.

The primary conclusion from our visible light data is that
the lack of a correlation between the line bisector span
and the RV is not proof of a reflex motion companion.  In addition, the
RV period can change significantly with new
data.  Initial analysis of 20 visible light RVs for DN Tau,
taken over 2.5 years, indicated convincing modulation with P$=$7.5 days and
FAP$=$0.002.  Although not markedly different from rotation period
estimates of 6.0$-$6.4 days \citep{bou93, per06}, $\chi^2$ for P$=$7.5 days
was more favorable than that for a secondary
peak at 6.3 days.  Removing the fit
to either the 7.5 or 6.3 day period from the RV data and recalculating
the power spectrum yielded no significant peak, suggesting that the
data were best represented with only one period.  With the
addition of RVs measured in winter 2007$-$2008, the integrity of the
7.5 day period was diminished and the 6.3 day period came to
dominate the power spectrum.
The rotation period of V836 Tau has been stable for decades at 6.76 days
\citep{gra08}.  We find a different RV period, 2.48 days, and 
substantially reduced power in the RV modulation near 6.76 days.
The RV period for V827 Tau, 3.76 days, is indistinguishable
from previously determined values of the rotation period \citep{gra08}.

If the visible light RV modulation originates in spots,
why don't the bisector spans correlate (Figures 1 and 2)?  Figure 3
clearly shows correlation for V827 Tau; many other stars
show a correlation as well \citep[e.g.,][]{que01, bou07, hue08}.
We do not believe that this can be attributed to a
stronger impact of spots on particular spectral lines because the same
echelle orders were used to determine the bisector spans for all targets.
\citet{des07} show that when $v$sin$i$ is smaller than the spectrometer
resolution, RV and bisector variations originating in spots can mimic the
behavior expected from short period giant planets.  The $v$sin$i$ of
DN Tau and V836 Tau, $\sim$10~km~s$^{-1}$,
are not much larger than our visible light spectral resolution, 5~km~s$^{-1}$,
while V827 Tau and LkCa 19 (Huerta et al. 2008) have $v$sin$i$ values of
$\sim$20~km~s$^{-1}$, suggesting qualitative agreement
with the simulations of \citet{des07}.

\section{Summary}

We have measured the RVs of the $\sim$2~Myr old T~Tauri stars DN~Tau, V836~Tau,
and V827~Tau in visible and IR light; the variations
we see in all three systems are likely the result of
starspots.  Furthermore:  (1) Periodogram analysis can
reveal the presence of a period but not necessarily a reliable value
until RV measurements densely sample the full phase of the periodic
signal.  (2)  The lack of correlation of
line bisector spans with RVs for some stars with spots
is an important problem to address in the search
for planets around young stars.  (3) High-resolution, IR spectroscopy is
critical for the verification of young planet candidates;
without the contrast in RV modulation amplitude
between the visible light and IR data, it is unclear whether
spots or a companion cause variability.
(4) In the case of the spotted young star, V827 Tau (Figure 3), we observe
a substantial reduction in RV amplitude between the visible
light and IR data.  This result can be exploited to improve our
understanding of starspot temperature and filling factors.
(5) T~Tauri stars are virtually guaranteed to have spots at some
level; if planets are also present, their detection will likely
require disentanglement of the blended RV signals and will benefit greatly
from measurements that span a broad range in wavelength.

\bigskip
\bigskip

The authors acknowledge the SIM Young Planets Key Project
(PI C. Beichman) for research support; funding was also
provided by NASA grant 05-SSO05-86.
We thank G. Blake, C. Salyk, and E. Schaller for sharing
Keck time (our first attempt at IR observations of DN Tau),
D. Gies and A. Fullerton for their assistance
implementing the CLEAN power spectrum estimate, and the
referee, J. Eisloeffel, for a prompt and helpful report.
LP thanks P. Bodenheimer and T. Barman for informative discussions.
This work made use of the SIMBAD database, the NASA
Astrophysics Data System, and the Two Micron All
Sky Survey, a joint project of the Univ. of Massachusetts
and IPAC/Caltech, funded by NASA and the NSF.
We recognize the significant cultural role that Mauna Kea
plays in the indigenous Hawaiian community and are
grateful for the opportunity to observe there.

\clearpage
\begin{figure}
\plotone{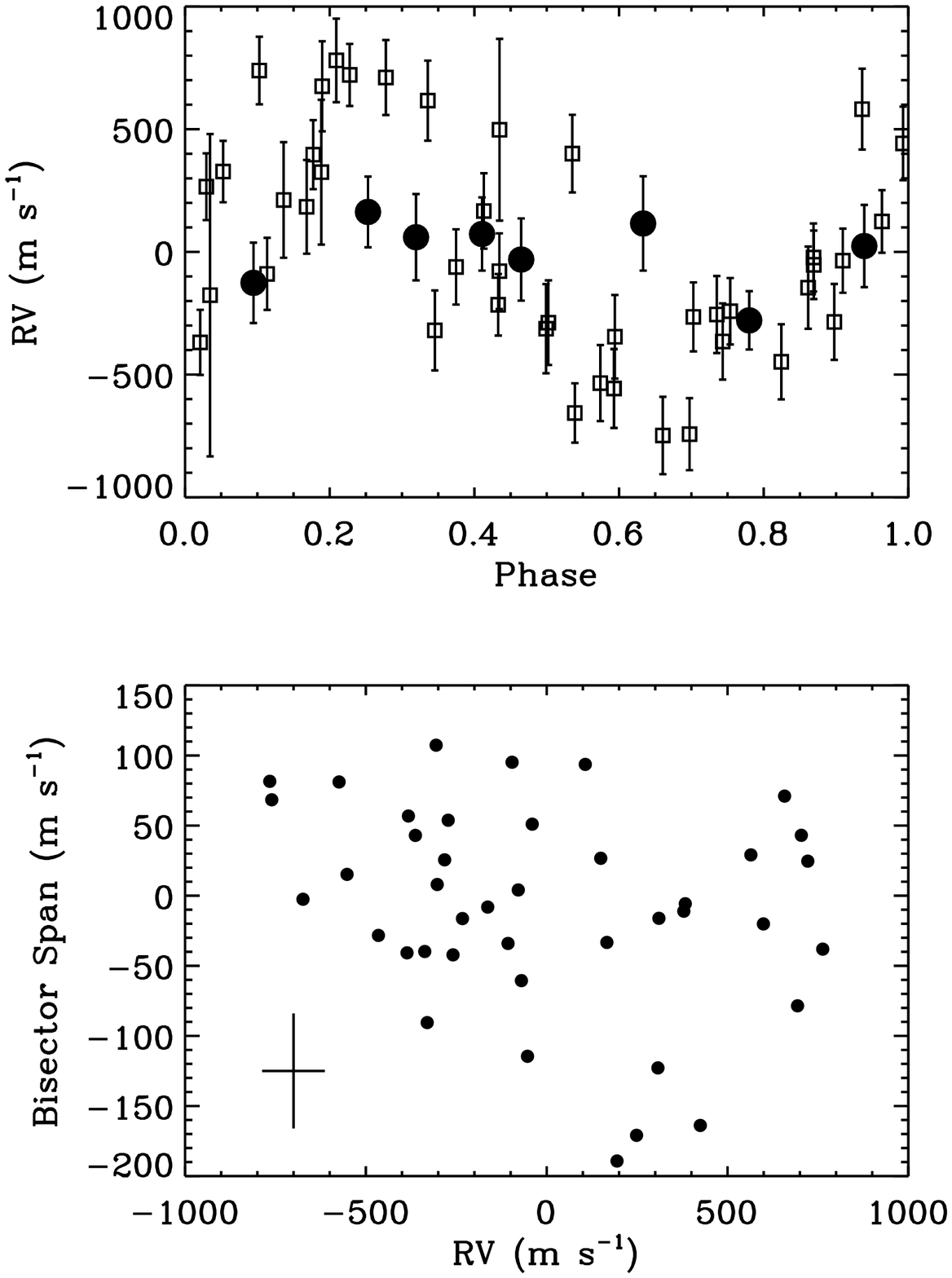}
\caption{{\it Top:} Relative RV versus phase for DN Tau.
Visible light (open squares) and IR (filled circles) data have been
phased to a priod of 6.33~$\pm$0.20 days.
{\it Bottom:} Line bisector spans as a function of RV for the visible
light data.  The linear correlation coefficient 
is $r = -0.27$ and the associated FAP, $f_p$, is 0.09.
\label{fig1}}
\end{figure}

\clearpage
\begin{figure}
\plotone{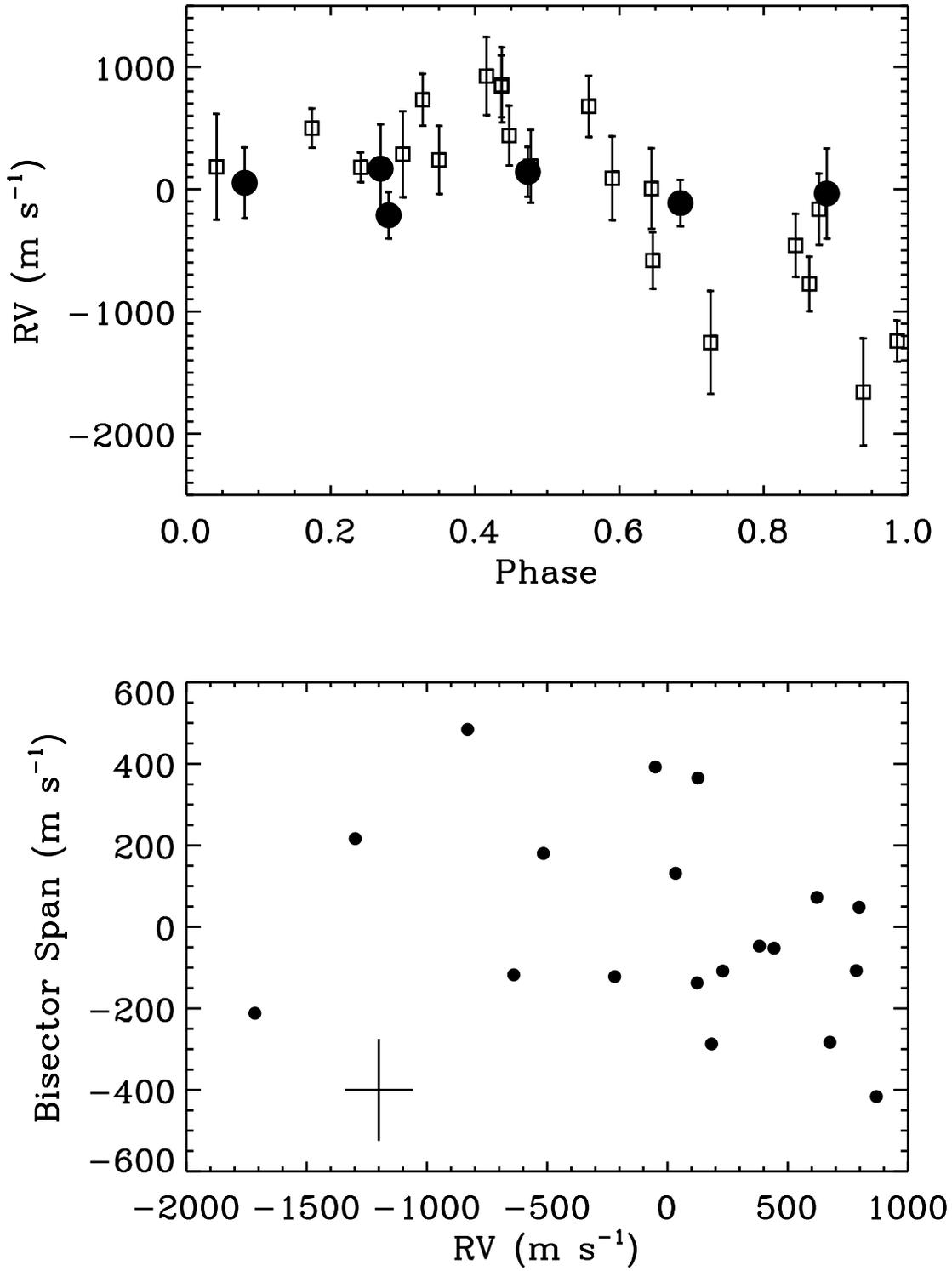}
\caption{Same as Figure 1 for V836 Tau, phased to a period of
2.48~$\pm$0.49 days ($r = -0.30$, $f_p = 0.20$).
\label{fig2}}
\end{figure}

\clearpage
\begin{figure}
\plotone{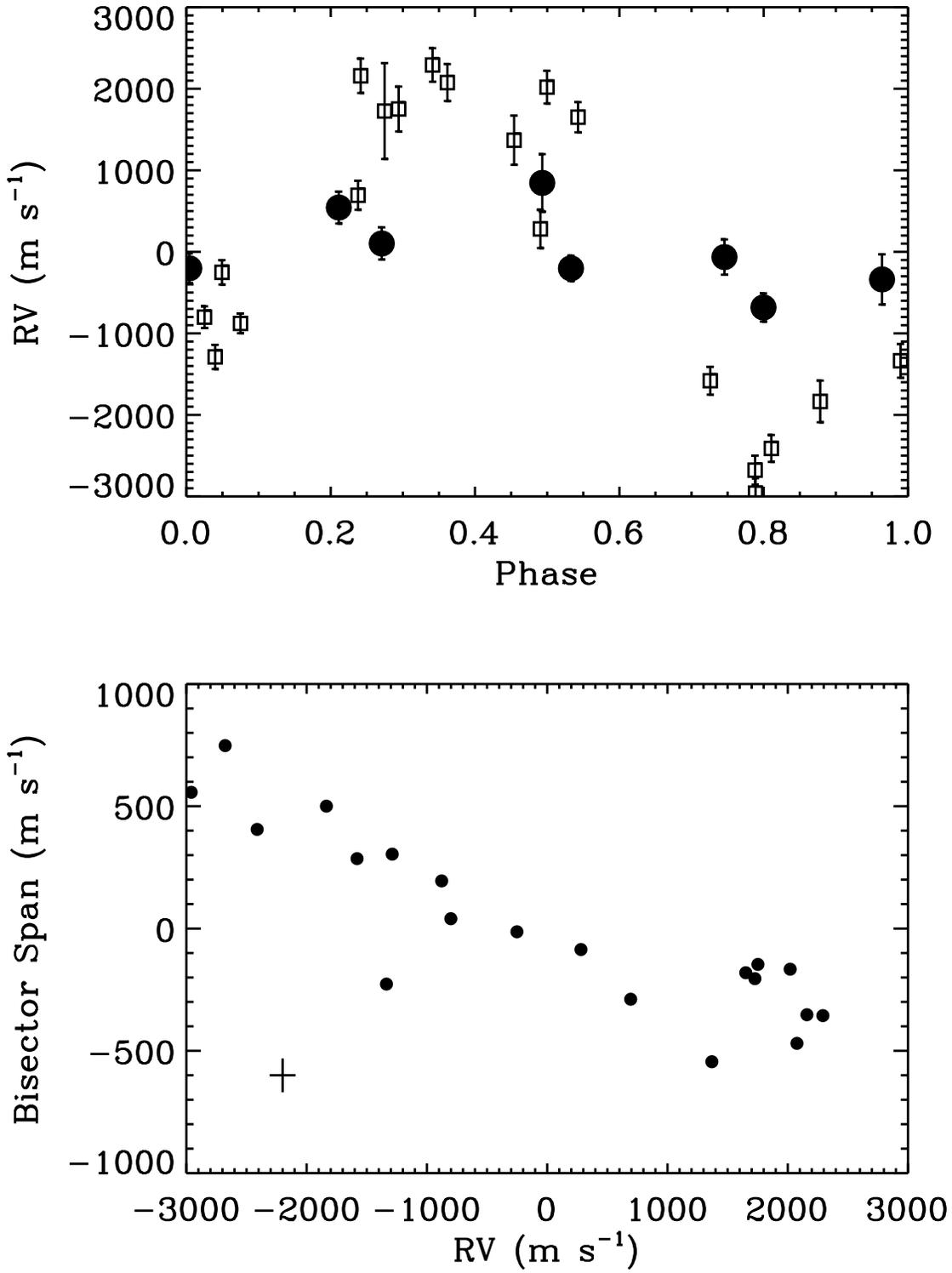}
\caption{Same as Figure 1 for V827 Tau, phased to a period of
3.76~$\pm$0.06 days ($r = -0.88$, $f_p = 2.8 \times 10^{-7}$).
\label{fig3}}
\end{figure}

\clearpage
\begin{figure}
\plotone{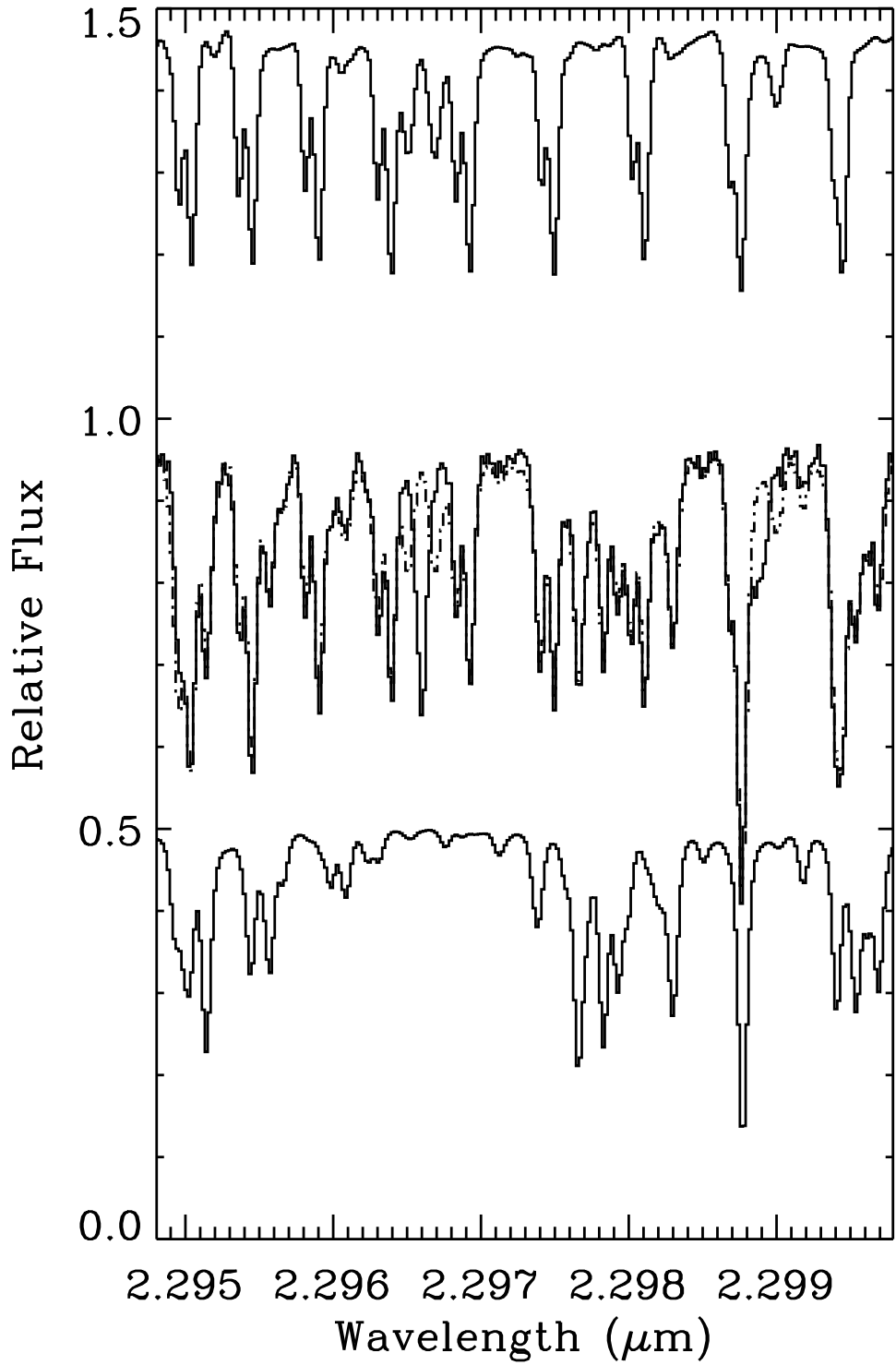}
\caption{{\it Middle:} UT 20 February 2008 CSHELL spectrum (solid line) of the RV
standard GJ~281 and a model (dash-dot line) consisting of a stellar photosphere
component ({\it Top:} from the NSO sunspot atlas) and a
telluric component ({\it Bottom:} from the NOAO spectral atlas).
\label{fig4}}
\end{figure}

\end{document}